\documentclass[prd,twocolumn,showpacs,groupedaddress,superscriptaddress,amsmath,amssymb]{revtex4-2} 
 
\newcommand{\vect}[1]{\boldsymbol{#1}}
\newcommand{\Avrg}[1]{\left\langle #1 \right\rangle}

\usepackage[utf8]{inputenc}
\usepackage[english]{babel}										
\usepackage{cmap}
\usepackage{bbm}
\usepackage[T2A]{fontenc}

\usepackage{enumerate}
\usepackage{multirow}
\usepackage{float}

\usepackage{xcolor}

\usepackage{amsmath,amsfonts,amssymb,mathtools}	
\usepackage{euscript} 
\usepackage{mathrsfs}
\usepackage{mathtext}

\usepackage{color}			
\usepackage{graphicx}
\usepackage[export]{adjustbox}
\DeclareGraphicsExtensions{.png,.jpg}
\graphicspath{}	
\usepackage[section]{placeins} 

\usepackage{physics}
\usepackage{braket}
\usepackage{xparse}	

\usepackage{slashed}
\usepackage{indentfirst} 
\usepackage{enumitem}

\usepackage[colorlinks=true, filecolor=blue, citecolor=blue,urlcolor=black]{hyperref}
\usepackage{url}
\usepackage{cleveref}
\date{\today}

\begin{document}

\title{Probing scalar, Dirac, Majorana and vector DM through  spin-0 electron-specific mediator at the electron fixed-target experiments} 

\author{I.~V.~Voronchikhin}
\email[\textbf{e-mail}: ]{i.v.voronchikhin@gmail.com}
\affiliation{ Tomsk Polytechnic University, 634050 Tomsk, Russia}

\author{D.~V.~Kirpichnikov}
\email[\textbf{e-mail}: ]{dmbrick@gmail.com}
\affiliation{Institute for Nuclear Research, 117312 Moscow, Russia}

\begin{abstract}
We discuss the thermal target curves of Majorana, Dirac, scalar and vector light dark 
matter (DM) that are associated with the freeze-out mechanism via the annihilation into $e^+e^-$ pair through
the electron-specific spin-0 mediator of dark matter. We also discuss the mechanism to produce 
the regarding  DM mediator in the electron (positron) fixed-target 
experiments such as NA64e and LDMX. We derive the corresponding
experimental reaches of the NA64e and LDMX that are complementary
to the DM thermal target parameter space.   
\end{abstract}

\maketitle

\section{Introduction}

A significant number of astrophysical observations lead to the hypothesis of dark matter (DM), 
that  manifests itself via the gravitational effects (for a review 
see e.~g.~Refs.~\cite{Bergstrom:2012fi,Bertone:2016nfn}). 
However, the nature of DM particles still remains of the great interest, despite the 
overwhelming evidences of indirect observation.
 Moreover, 85~$\%$ of the total amount of
 matter~\cite{Planck:2018vyg} in the Universe is approximately invisible and  that motivates a development of extension of the Standard model (SM).
In addition, DM particles  can solve other well-known problems of modern physics, for instance  
the anomalous magnetic moment puzzle~\cite{AOYAMA20201}.

The gravitational evidences cannot directly specify 
the range of DM particle masses that can span a significant  orders of magnitude leading to different cosmological consequences. 
However, if DM particles achieve thermalization with SM particles in the thermal bath of the early Universe,  the range of viable masses decreases.
In particular, DM masses below a few $\mbox{keV}$ are too hot for structure formation and DM 
masses above $100\;\mbox{TeV}$ are in tension with perturbative 
unitarity~\cite{Krnjaic:2015mbs}.
The expansion of the Universe results in DM freeze-out that leads to its decoupling  and 
the relic abundance.
It is worth mentioning that thermal freeze-out is relatively insensitive to initial 
conditions of the early Universe~\cite{Hall:2009bx,Brahma:2023psr}.

Thermal contact between DM and SM particles lead to overproduction of DM particles in the early Universe, therefore the relic abundance requires a depletion mechanism to yield the observed value~\cite{Krnjaic:2015mbs}.
Further, the assumption of the weak interaction between DM and SM particles lead to weak interacting massive particle (WIMP) that takes masses in the Lee-Weinberg bounds~\cite{Lee:1977ua,Kolb:1985nn} from few $\mbox{GeV}$ to $100\;\mbox{TeV}$. 
However, the heavy range of WIMP masses is ruled out by the direct detection experiments  ~\cite{LUX:2016ggv,XENON:2017vdw,PandaX-II:2017hlx} that motivates the search of light DM in the sub-GeV mass region. 
To avoid the DM overproduction one can suggest the scenario with the mediator of DM (MED) that can connect light DM and SM particles via portals.

In particular, the typical scenarios and the regarding thermal target associated with  
dark boson mediators include spin-0, spin-1 and spin-2 particles such as the hidden Higgs boson~\cite{McDonald:1993ex,Burgess:2000yq,Wells:2008xg,Schabinger:2005ei,Bickendorf:2022buy,Boos:2022gtt,Sieber:2023nkq}, the dark photon~\cite{Catena:2023use,Holdom:1985ag,Izaguirre:2015yja, Essig:2010xa, Kahn:2014sra,Batell:2014mga,Batell:2017kty,Izaguirre:2013uxa,Kachanovich:2021eqa,Lyubovitskij:2022hna,Gorbunov:2022dgw,Claude:2022rho,Wang:2023wrx},  and   the dark  graviton \cite{lee:2014GMDM,Kang:2020-LGMDM,Bernal:2018qlk,Folgado:2019gie,Kang:2020yul,Dutra:2019xet,Clery:2022wib,Lee:2014caa,Gill:2023kyz,Wang:2019jtk,deGiorgi:2021xvm,deGiorgi:2022yha,Jodlowski:2023yne}, respectively. 
Also, in the mass range of light DM from few MeV to GeV the fixed-target experiments combine the advantages of high-energy particle beam and its relative large intensity.  


In this paper, we focus on DM in the mass range between $\mathcal{O}(1)$~MeV  and 
$\mathcal{O}(1)$~GeV. 
In the following we consider the DM parameter space with a typical branching fraction of the 
scalar electron-specific mediator  $\text{Br}({\rm MED \to DM\;DM}) \simeq  1$, that implies  the condition for the masses  $2 m_{\rm DM} \lesssim  m_{\rm MED}$.
We consider several simplified scenarios with specific DM types such as Majorana, Dirac, scalar and  vector particles. 
Moreover, we use effective interactions of electron-specific scalar MED that can be potentially coupled  with SM particles via singlet extension of the Higgs sector. 
We calculate a new constraint on the coupling between the scalar MED and a electron by using 
the null-results of fixed-target experiments such as NA64e and LDMX. We also show the 
complementarity of  the derived bounds  to the thermal target curves of the specific DM type.
We pay additional attention to the resonant annihilation of electron-positron pair into DM via the scalar MED that also provides a new constraint on the light DM parameter space from NA64e and LDMX.

This paper is organized as follows. 
In Sec.~\ref{sec:BenchModels} we describe the benchmark scenarios for the electron-specific 
scalar MED.   In Sec.~\ref{sec:RelicAbundance} we briefly provide the procedure for the 
calculation of relic  density of DM for the simplified models.
In Sec.~\ref{sec:ExperimentalBenchmark} we discuss the missing energy signal and review the 
main benchmark parameters of  electron fixed-target facilities.
In Sec.~\ref{sec:ProductionMechanism}  we derive explicitly the decay widths of scalar mediator 
and the resonant cross sections of the process $e^+ e^- \to \phi \to \mbox{DM} + \mbox{DM}$ at 
NA64e and LDMX.   In Sec.~\ref{sec:FreezeOutMechanism} we obtain  the thermal target curves of 
various types of DM. In Sec.~\ref{sec:ExpectedReach} we discuss the experimental reach of  NA64e 
and  LDMX for the the specific light DM parameter space.  
We conclude in Sec.~\ref{sec:Conclusion}.

\section{Benchmark scenarios
\label{sec:BenchModels}}

In this section we  discuss simplified benchmark DM scenarios 
focusing on the electron-specific scalar mediator interacting with scalar, Dirac, Majorana, and vector DM. 
By diagonalizing the mass terms  after the electroweak symmetry breaking  in the effective field 
theory~\cite{Djouadi:2011aa,Lebedev:2011iq,Krnjaic:2015mbs} one can obtain the  
coupling  between SM particles and the scalar MED. 

To be more concrete we consider four types of the effective Lagrangians involving scalar MED $\phi$, electron $e$ and  the 
specific types of DM 
\begin{align}
    &   \mathcal{L}_{\rm eff}^{\rm S} 
\supset  
    \frac{1}{2} c^{\phi}_{SS} \phi S^2 + c^{\phi}_{ee} \phi \overline{e}  e,
\\
    &       \mathcal{L}_{\rm eff}^{\psi} 
\supset 
    c^{\phi}_{\psi\psi} \phi \overline{\psi} \psi + c^{\phi}_{ee} \phi \overline{e}  e,
\\
    &   \mathcal{L}_{\rm eff}^{\chi} 
\supset  
    \frac{1}{2} c^{\phi}_{\chi \chi} \phi \overline{\chi} \chi + c^{\phi}_{ee} \phi \overline{e}  e,
\\
    &   \mathcal{L}_{\rm eff}^{\rm V} 
\supset  
    \frac{1}{4} c^{\phi}_{VV} \phi V^{\mu \nu} \widetilde{V}_{\mu \nu} 
    +
    c^{\phi}_{ee} \phi \overline{e}  e,
\end{align}
where $c^{\phi}_{SS}$, $c^{\phi}_{\psi\psi}$, $c^{\phi}_{\chi \chi}$, $c^{\phi}_{VV}$ are coupling 
constants for  real scalar DM $S$~\cite{Djouadi:2011aa}, Dirac fermion DM 
$\psi$~\cite{Krnjaic:2015mbs,Marsicano:2018vin},  Majorana  fermion $\chi$~\cite{Djouadi:2011aa,Berlin:2018bsc} and vector DM
$V_\mu$~\cite{Kaneta:2017wfh,Nomura:2008ru,Kaneta:2016wvf,Zhevlakov:2022vio}  with the spin-0 mediator, respectively, $c^{\phi}_{ee}$ 
is the constant of coupling between the scalar MED and the electron~\cite{Batell:2017kty,Berlin:2018bsc}, $V^{\mu \nu}$ is the field strength tensor,  
$\widetilde{V}_{\mu \nu} = 1/2 \; \epsilon^{\mu \nu \alpha \beta} V_{\alpha \beta}$ is the dual field strength tensor.

In is worth noticing that for the Majorana DM we employ the 4-component spinor 
formalism~\cite{Dreiner:2008tw}. We introduce an ubiquitous in a literature notations for the 
couplings  
\begin{equation}\label{eq:forCouplingConstant}
    ( \epsilon^{\phi}_{ee} )^2 = ( c^{\phi}_{ee} )^2 / ( 4 \pi \alpha ), 
\quad
    \alpha_{\rm DM} = (c_{ \text{\rm DM\; DM}}^\phi )^2/(4 \pi),
\end{equation} 
where $\alpha_{\rm DM}$ and $\alpha \approx 1/137$ are the dark and electromagnetic fine structure 
constants, respectively. For concreteness we also assume throughout the paper that 
the visible decays of mediator $\phi \to e^+ e^-$ are subdominant with respect to the invisible  $\phi \to \mbox{DM} + \mbox{DM}$, such that its DM branching fraction  is dominant, $\mbox{Br} (\phi \to \mbox{DM} + \mbox{DM})\simeq 1$.  

\section{Relic abundance}\label{sec:RelicAbundance}

In this section we discuss in detail the procedure of calculation for the relic abundance of DM due to the 
annihilation freeze-out reaction  
 $p_1^{\rm DM} + p_2^{\rm DM} \to p_a^{\rm SM} + p_b^{\rm SM}$. The Boltzman equation  
 can be written by defining the term 
$Y(x) = n(x)/s(x)$ as a ratio of DM number density $n(x)$ over the SM entropy 
density $s(x)$ in the following form:
\begin{equation}\label{eq:BoltzmannEq2to2}
    \frac{d Y}{d x} 
\! = \!
    - Q(x) 
    \left( Y^2(x) - Y^2_{\rm eq}(x)\right),
\;\;
    Y_{\rm eq}(x) 
\! = \! 
    \frac{n_{\rm eq}(x)}{s(x)},
\end{equation}
where $x = m_{\rm DM} / T$, $T$ is the typical temperature, 
 $n_{\rm eq}(x)$ is the density of particle number in the thermal equilibrium. The function $Q(x)$ and the total entropy density $s$ read, respectively:

\begin{equation}
Q(x)
\! = \!
   \frac{c \; s(m_{\rm DM}) }{H(m_{\rm DM})} \Avrg{\sigma v_{\rm Mol}},
\quad
    s(T) 
\! = \!
    \frac{2 \pi^2}{45} 
    g_{*s} \!\! \left(T\right) 
    T^3,
    \label{QandSdef1}
\end{equation}
where $g_{*s}(T)$ is the entropy effective degree of freedom \cite{Husdal:2016haj}. The coefficient is chosen to be $c = 1$ for the 
case when  DM is its own antiparticle (Majorana fermion, scalar and  vector types of DM), and $c = 1/2$ for the 
opposite case  (i.~e.~ for the Dirac fermion DM) \cite{Srednicki:1988ce}. The Hubble parameter is defined as \cite{Kolb:1990vq}:
\[
    H(x) 
 \! \simeq  \! 
    \frac{H(m_{\rm DM})}{x^2},
\quad
    H(m_{\rm DM}) 
 \! \simeq  \!
    0.331 \;
    g_{*\varepsilon}^{1/2}(m_{\rm DM})
    \frac{m_{\rm DM}^2}{M_{Pl}},
\]
where $M_{Pl} \simeq  2.4 \cdot 10^{18} \;\mbox{GeV}$ is the reduced Planck mass, $g_{*\varepsilon}(T)$ is the energy 
effective degree of freedom \cite{Husdal:2016haj}, $\Avrg{\sigma v_{\rm Mol}}$ is the thermal averaged  
cross section for a reaction $2 \to 2$ that is given by~\cite{Gondolo:1990dk}:
\begin{equation}
\label{eq:ThermalAverege2to2}
    \Avrg{\sigma v_{\rm Mol}} 
=
    \frac
    {\int \sigma v_{\rm Mol} e^{-E_1/T} e^{-E_2/T} d^3 \vect{p}_1  d^3 \vect{p}_2 }
    {\int e^{-E_1/T} e^{-E_2/T} d^3 \vect{p}_1  d^3 \vect{p}_2 },
\end{equation}
where $p_1^{\rm DM} = (E_1, \vect{p}_1), p_2^{\rm DM} = (E_2, \vect{p}_2)$ are 4-momenta of DM, $v_{\rm Mol}$ is the Moller velocity 
\cite{Cannoni:2016hro}:
\begin{equation}\label{eq:MollerVelocityGeneral}
    v_{\rm Mol} 
= 
    \frac{\sqrt{\lambda(s, m_{\rm DM}^2,m_{\rm DM}^2)}}{2 E_1 E_2},
\end{equation}
and $\lambda(s, m_1^2,m_2^2)~=~(s - (m_1 + m_2)^2)(s - (m_1 - m_2)^2)$ is the triangle 
function. The expressions for the total cross section of the DM annihilation $\sigma$  are  
provided explicitly in Sec.~\ref{sec:FreezeOutMechanism} for the specific DM type. In the
expression of thermal average of cross section \eqref{eq:ThermalAverege2to2} we neglect 
 the chemical  potential and assume the Maxwell-Boltzmann distribution for the particle 
 species.  

\begin{figure*}[!ht]
\centering
\includegraphics[width=1\textwidth]{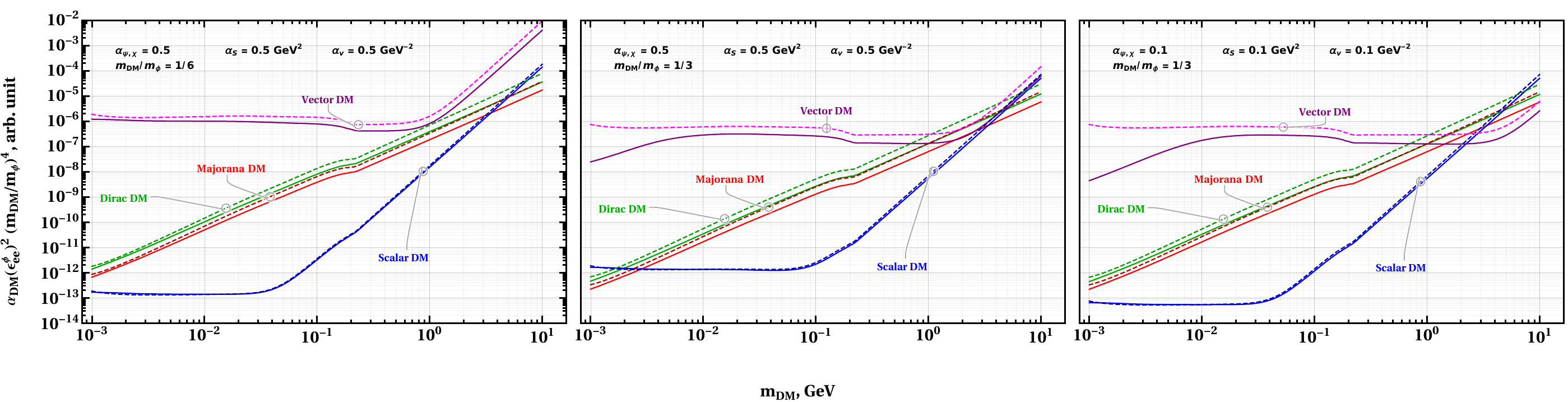}
\caption{ 
The relic DM thermal target curves as a function of mass for Majorana DM (red line), Dirac DM (green line), vector DM (purple line) and scalar DM (blue line) that annihilate via scalar MED. 
Solid lines  correspond to the numerical calculation through  integral \eqref{eq:GelminiTermAvCS} and the dashed lines are  associated with semi-analytical calculation of the thermal target in the  low-velocity approach~\eqref{eq:LowVelocityExpan}. 
 Left panel: relic thermal targets for $\alpha_{\rm DM} = 0.5\, (\mbox{GeV}^{2}/\mbox{GeV}^{-2})$ and $m_{\rm DM} / m_{\phi} = 1/6$. 
Center panel: relic targets for $\alpha_{\rm DM} = 0.5 \, (\mbox{GeV}^{2}/\mbox{GeV}^{-2})$ and  $m_{\rm DM} / m_{\phi} = 1/3$. 
The right panel: relic targets for $\alpha_{\rm DM} = 0.1 \, (\mbox{GeV}^{2}/\mbox{GeV}^{-2})$ and $m_{\rm DM} / m_{\phi} = 1/3$.
}\label{fig:RelAbAll}
\end{figure*}

The decoupling of the considered type of particles is achieved if the following condition is 
fulfilled:
\begin{equation}\label{DecouplingCondition}
     c\; n_{\rm eq}(x_f) 
    \left. \Avrg{\sigma v_{\rm Mol}}  \right|_{x = x_f}
\simeq   
    H(x_f),
\end{equation}
where $T_f$ is the critical temperature  and $x_f = m_{\rm DM} / T_f$ is the parameter of freeze-out. 
 The current value of variable $Y$ tends to the relic value $Y(\infty)$, that can be estimated on the interval $(x_f, \infty)$ from the Boltzmann equation \eqref{eq:BoltzmannEq2to2}~as:
\begin{equation}\label{eq:exprYinfty}
    Y^{-1}\!(\infty) 
\! = \!
    \frac{c \; s(m_{\rm DM}) }{H(m_{\rm DM})} 
    \! J(x_f),
\;\;
    J(x_f)
\! = \!
    \! \int\limits_{x_f}^{\infty} \!
    \frac{\Avrg{\sigma v_{\rm Mol}}}{x^2} dx,
\end{equation}
where we take into account that $Y(x)~\gg~Y_{\rm eq}(x)$ for $x~>~x_f$. 
In case of s-wave annihilation one can exploit $J(x_f) = \Avrg{\sigma v_{\rm Mol}} / x_f$.
Next, from the Friedmann equations for the critical density of the DM one can obtain that
\begin{equation}\label{eq:AproxRelicDensityOneType}
    \Omega_{\rm DM} 
\simeq 
    \frac{ m_{\rm DM} s_0 Y(\infty)}{3 M_{Pl}^2 H_0^2},
\end{equation}
where  $s_0 = s(T_0)$, $T_0 = 2.35 \cdot 10^{-13} \;\mbox{GeV}$ are the current total entropy density and temperature of 
the Universe, $H_0~=~2.13~h~\cdot~10^{-42}~\mbox{GeV}$ is the current value of the Hubble 
parameter with the dimensionless constant $h \simeq 0.674 \pm 0.005$~
\cite{Planck:2018vyg}. As a result,  the expression of relic density takes the following form \cite{Kolb:1990vq,Kahn:2018cqs,Kang:2020-LGMDM}:
\begin{equation}\label{eq:AproxRelicDensityOneTypeResult}
    \Omega_{c} h^2
\!  \simeq  \! 
    0.85  \! \cdot \! 10^{-10} 
    g^{-1/2}_{*s}(m_{\rm DM}) \frac{1}{c}
    \left(\frac{ J(x_f) }{ \mbox{GeV}^{-2} }\right)^{-1},
\end{equation}
where it is taken into account that $g_{*s}(T_0) \approx 3.91$ and for $T \gtrsim 1 \;\mbox{MeV}$ we can set $g_{*\varepsilon}(T) \simeq  g_{*s}(T)$. 
In addition, 
we take into account that the current value of relic abundance of cold DM obtained from 
the Planck 2018 combined analysis is~\cite{Planck:2018vyg}:
\[
    \Omega_c h^2 = 0.1200 \pm 0.0012.
\]
Moreover, assuming s-wave annihilation one can see from the relic density \eqref{eq:AproxRelicDensityOneTypeResult} that the typical value of thermal averaged  
cross section is evaluated as $\Avrg{\sigma v_{\rm Mol}}~\simeq~2.5~\times~10^{-9}~\mbox{GeV}^{-2}~\approx~3~\times~10^{-26}~\mbox{cm}^{-3}~s^{-1}$~\cite{Lee:2014caa,Cai:2021nmk} where we imply that $c = 1$ and the typical parameters are  $m_{\rm DM} \simeq \mathcal{O}(10)\;\mbox{MeV}$,  $g_{*s}~\simeq~\mathcal{O}(10)$ and  $x_f \simeq \mathcal{O}(10)$. 
Next, using the direct integration of expression \eqref{eq:ThermalAverege2to2}, the thermal 
averaged cross section for $x \gtrsim 1/3$ can be written
as~\cite{Gondolo:1990dk,Bharucha:2022lty,Krnjaic:2015mbs}: 
\begin{align}\label{eq:GelminiTermAvCS}
     &  \Avrg{\sigma v_{\rm Mol}} \!\!
\simeq  \!\!
    \! \! 
    \int\limits_{4 m_{ \rm DM }^2 }^{\infty} 
    \! \! \! \! \! 
    \sigma(s) 
    \frac{\lambda(s, m_{ \rm DM }^2,m_{ \rm DM }^2)}{N_{\Avrg{\sigma v_{\rm Mol}}} \sqrt{s}}
    K_1 \! \! 
    \left(\frac{\sqrt{s}}{T}\right) ds,
\\
    &   N_{\Avrg{\sigma v_{\rm Mol}}} 
= 
    8 T  m_{ \rm DM }^4  \left(K_2\left(m_{ \rm DM }/T\right) \right)^2,
\end{align}
where $K_i(z)$ is the modified Bessel functions of the second kind of $i^{\text{th}}$ 
order~\cite{abramowitz1988handbook}.  
Also,  we use the critical temperature $T_f = m_{ \rm DM } / 20$ \cite{Lee:1977ua,Krnjaic:2015mbs} for the 
calculation of relic abundance  with the integral \eqref{eq:GelminiTermAvCS}.

\subsection{Low-velocity approach}

 By using the velocity expansion for the thermal averaged of cross section in 
the non-relativistic approach 
one can 
get~\cite{Gondolo:1990dk,Choi:2017mkk}:
\begin{multline}\label{eq:LowVelocityExpan}
    \Avrg{\sigma v_{\rm Mol}}
=
    \frac{1}{2 \sqrt{\pi}}
    \sum\limits_{k=0}^{\infty}
    4^{k + 1}
    \Gamma(k + 3/2)
\;
    \frac{a_k}{k!}
    x^{-k}
\approx \\  \approx
    a_0 + 6 a_1 x^{-1} + 30 a_2 x^{-2},
\end{multline}
where $\Gamma(k)$ is the Gamma function, $a_k$ are the expansion coefficients of cross section for the low-velocity series.
Indeed, the low-velocity approach is possible due to the angular independence of Mandelstam variables  in the zero velocity~\cite{Wells:1994qy}.
As a result, for the calculation of relic density one can use a first non-zero term in the low-velocity approach as $\Avrg{\sigma v_{\rm Mol}} = \sigma_0 x^{-n}$. 
It is worth mentioning that $n = 0$, $n = 1$ and $n = 2$  correspond to the  s-wave, p-wave and 
d-wave annihilations, respectively \cite{Kolb:1990vq}.
Hence, by taking into account the condition of decoupling \eqref{DecouplingCondition} and 
the relic density \eqref{eq:AproxRelicDensityOneType} one can get final
expressions for a calculation of relic abundance in the low-velocity approach~\cite{Kolb:1990vq,Kahn:2018cqs}:
\begin{align}
    &   \Avrg{\sigma v_{\rm Mol}} = (c^{\phi}_{\rm ee} )^2 \sigma_0 x^{-n},
\\
    &   x_f \label{eq:xfLowVelocity}
= 
    \ln
    \left(
        (c^{\phi}_{\rm ee} )^2
        \sigma_0
        \frac{ 3\sqrt{5}M_{Pl} }{ 2 {\pi}^{5/2} }
        \frac{g_{i} m_{\rm DM} }{\sqrt{g_{*s}(m_{\rm DM})} }
    \right),
\\
    &   \Omega_{c} h^2 \label{eq:RelicDensityLowVelocity}
\! = \!
    \frac{0.85 \cdot 10^{-10}}{ c ( c^{\phi}_{\rm ee} )^2 }  
    \frac{(n \! + \! 1) x_f^{n + 1} }{ g^{1/2}_{*s}(m_{\rm DM}) \sigma_0}
    \;\mbox{GeV}^{-2},
\end{align}
where $g_{i}$ - the internal degrees of freedom, $n$ is the first non-zero power in the low-velocity series. 
 
 The calculation of relic density is associated  with the Eqs.~(\ref{eq:xfLowVelocity}) 
and~(\ref{eq:RelicDensityLowVelocity}) for $c^{\phi}_{\rm ee}$ and $x_f$ in case of low-velocity 
approximation. Opposite, we have the direct expression for parameters $c^{\phi}_{\rm ee}$ with $x_f = const$ 
in case using the integral formula for the thermal average of cross section \eqref{eq:GelminiTermAvCS}.
Indeed, in case of low-velocity approximation the relic density depends polynomially on the freeze-out parameter and this system can be analytically solved for $x_f$ and $c^{\phi}_{\rm ee}$. 

\section{Missing energy signal \label{sec:ExperimentalBenchmark}}

In this section we  discuss the electron  missing energy signatures and the typical  parameters of the NA64e and LDMX experiments for probing DM mediators.

We estimate the number of MEDs produced due to the bremsstrahlung for fixed-target facilities as follows:  
 \begin{equation}\label{eq:NradGrav}
N^{\rm brem. }_{\rm MED} \simeq \mbox{EOT}\cdot \frac{\rho N_A}{A} L_T \int\limits^{x_{max}}_{x_{min}}
dx \frac{d \sigma_{2\to3}(E_0)}{dx}\eta_{\rm MED}^{\rm brem.},
\end{equation}
where   $L_T$ is the effective interaction length of the electron in the target,  $\mbox{EOT}$ is a number of  electrons accumulated on target, $\rho$ is the target density, $N_A$ is Avogadro's number,  $A$ is the atomic weight number, $Z$ is the atomic number, $\eta_{\rm MED}^{\rm brem.}$ is a typical efficiency for the bremsstrahlung emission of the MED, $d\sigma_{2\to3}/dx$ is the differential cross section of the electron missing energy process $e N \to e N \phi$ \cite{Voronchikhin:2023znz,Liu:2016mqv}, $E_0 \equiv E_{beam}$ is the initial energy of electron beam,   $x_{min}$ and $x_{max}$ are the minimal and maximal fraction of missing energy, respectively, for the experimental setup, $x\equiv E_{miss}/E_0$, where $E_{miss} \equiv  
E_{\rm MED}$.

Also, the typical number of hidden bosons produced due to the annihilation is estimated to 
be:
\begin{equation}\label{eq:constrCoulConst}
    N^{\rm ann. }_{\mbox{\scriptsize MED}}\! \simeq \! \mbox{EOT} \frac{\rho N_A \! Z \! L_T }{A}\! \! \!\!\!\! \int\limits_{E_{e^+}^{cut}}^{E_{e^+}^{max}} \!
    \!\!\!\!\!  dE_{e^+} \! \sigma_{tot} (\!E_{e^+}\!)\! T(\!E_{e^+}\!)\eta_{\rm MED}^{\rm ann.},
\end{equation}
where  $\sigma_{tot} (E_{e^+})$ is the resonant total cross section of the electron-positron 
annihilation into DM (see  Sec.~\ref{sec:ProductionMechanism} below), $\eta_{\rm MED}^{ann.}$ is a typical 
efficiency associated with MED production via the resonant channel (we conservatively imply throughout 
the paper that signals for both positron and electron beam modes have the same efficiency 
$\eta_{\rm MED}^{\rm ann.}$), $E_{e^+}$ is the energy of secondary positrons,  
$E_{e^+}^{cut}=E_{0} x_{min}$ and $E_{e^+}^{max} = E_{0}$ are the minimal and maximal energies of 
secondary positrons in the electromagnetic shower, respectively.

\subsection{Positron track-length distribution
\label{sec:PosTrLgDis}}

In this subsection we briefly discuss the using positron track-length distribution in a target  due to the electromagnetic shower development from the incoming  primary electron or positron beams.

The analytical approximation for the typical differential positron track-length distribution  $T(E_{e^+})$ was studied in detail~\cite{Bethe:1934za,Carlson:1937zz,Landau:1938qvy,Tsai:1966js,Marsicano:2018krp}. 
For the thick target it was shown that $T(E_{e^+})$ depends, at first order, on a specific type of target material through the multiplicative factor $T(E_{e^+}) \propto X_0$, where $X_0$ is the radiation length. 
Moreover,  $T(E_{e^+})$ depends also on the ratio $E_{e^+}/E_0$, where $E_{0}$ is the energy of the primary impinging particle. 
This allows to exclude the dependence on the energy of the primary beam and radiation length. 
As a result, we can get the universal distribution of 
positrons~\cite{Marsicano:2018krp,Andreev:2021fzd,Voronchikhin:2023znz} adapted from 
Refs.~\cite{NA64:2022rme,Andreev:2021fzd} that is characterized  by the energy of primary beam $E_0$ 
and the target material~$X_0$ of the NA64e experiment. 
For the electromagnetic shower development of  positrons we use  
numerical Monte Carlo simulations in GEANT4~\cite{GEANT4:2002zbu}.
It is important to mention that the dependence of typical angles between the primary beam direction and 
a momentum of the secondary positrons can be neglected~\cite{Marsicano:2018krp}.

\subsection{Fixed-target experimental setup
\label{sec:ExperimentSetup}}

In this subsection we briefly summarize the benchmark parameters of the NA64e and LDMX electron fixed-target experiments 
that are exploited to constrain the parameter space of DM.

In case of fixed-target experiments with the initial energy of electron beam $E_0$, the fraction of the primary beam 
energy $E_{miss}=x E_0$ can be  carried away by a DM pair, that passes the detector of experiment without 
the energy deposition. The remaining part of the beam energy fraction, $E_e^{rec} \simeq (1-x) E_0$, can be 
deposited in the  active thick target by the recoil electrons (positrons). The fixed-target experiments 
allow investigating the relic DM with the mass in the range between 
$1\, \mbox{MeV}$ and $1\, \mbox{GeV}$. It is worth mentioning that both NA64$e$ and LDMX experiments have 
background suppression at the level of $\lesssim~ \mathcal{O}(10^{-13}-10^{-12})$.

{\it NA64e: } 
the NA64e is the fixed-target experiment located at CERN North Area with a beam from the Super Proton Synchrotron (SPS) H4 beamline.
The ultrarelativistic electrons with the  energy of $E_0 \simeq 100\, \mbox{GeV}$ can be exploited as the primary beam that is  scattering off nuclei of an active thick target.  
The typical scheme of the NA64e setup, the detector equipment and event selection rules  can be found elsewhere in Ref.~\cite{NA64:2022rme,NA64:2022yly,NA64:2023wbi}. 
The efficiencies $\eta_{\rm MED}^{\rm brem.}$ and $\eta_{\rm MED}^{ann.}$ are taken to be at the level of $90\%$ for both electron and positron beam modes~\cite{NA64:2022rme}.
In case of NA64e experiment we use the following benchmark parameters $(\rho~\simeq~11.34~~\mbox{g cm}^{-3}$, $A=207~~\mbox{g mole}^{-1}$, $Z=82$, $X_0=2.56~~\mbox{cm})$, the effective interaction length 
of the electron is $L_T=X_0$ and the missing energy fraction cut is $x_{min}=0.5$.
In this work we also perform the analysis of the sensitivity of NA64$e$ to probe DM for the accumulated 
data at the level of~$\mbox{EOT} \simeq 9.37 \times 10^{11}$ (see e.~g.~Ref.~\cite{NA64:2023wbi} for 
detail).

\begin{figure*}[!ht]
\centering
\includegraphics[width=1.\textwidth]{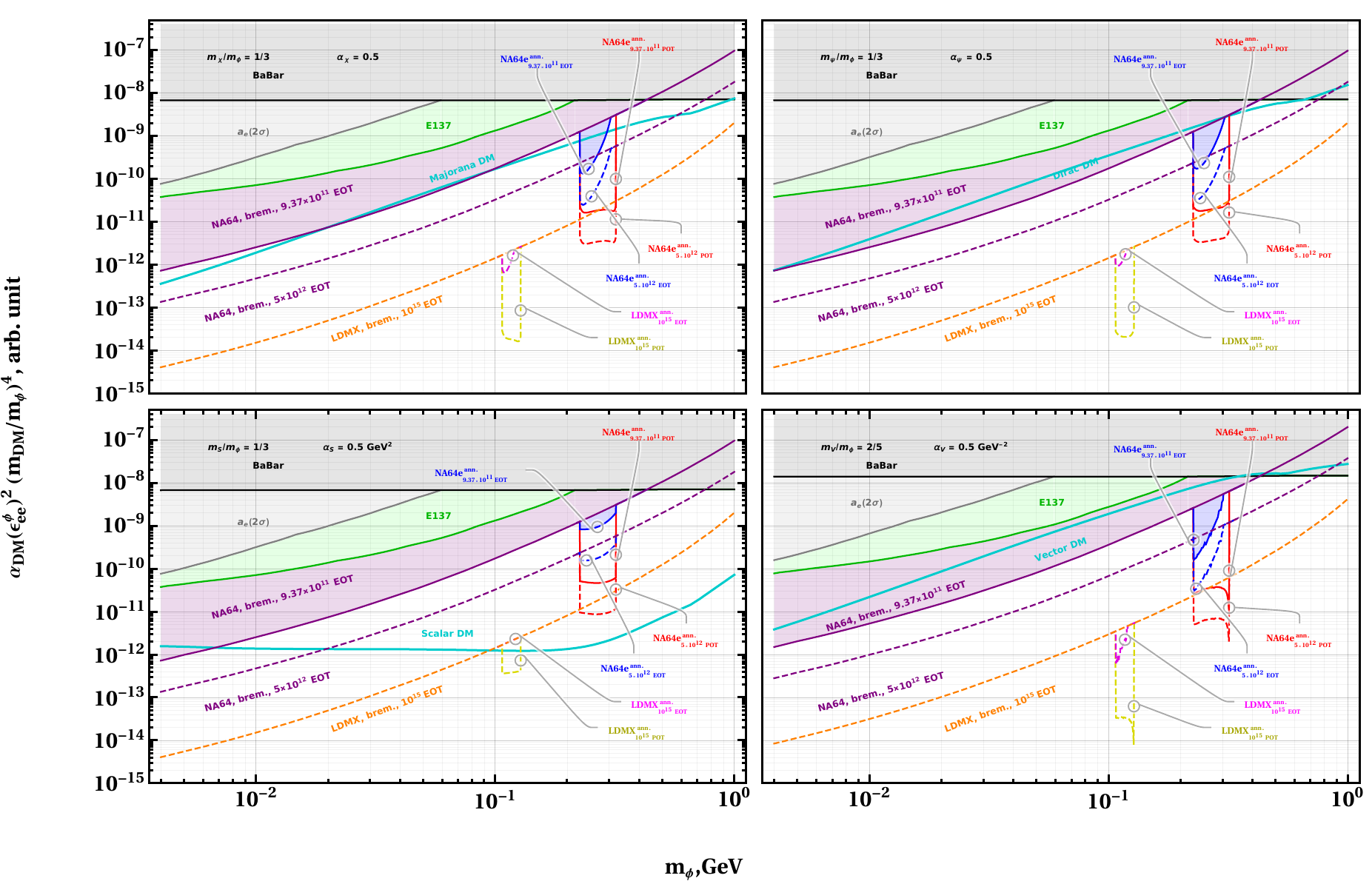}
\caption{ 
The experimental reach at $90\, \%$ C.L. as the function of MED mass for the NA64e and LDMX fixed-target facilities due 
to the $\phi$-strahlung  and resonant positron annihilation $e^+e^- \to \phi$, followed by the invisible decay  $\phi \to 
\chi \chi$.
For the calculation of relic thermal target we use numerical integration of \eqref{eq:GelminiTermAvCS}.
Grey shaded region corresponds to the exisiting BaBar~\cite{BaBar:2017tiz} monophoton limit $e^+e^- \to \gamma \phi$.
Green shaded region is the current reach of the  electron beam dump E137~\cite{Berlin:2018bsc,Bjorken:1988as,Batell:2014mga} experiment.
The solid purple line corresponds to the existing NA64e limits for $\mbox{EOT}\simeq 9.37\times 10^{11}$ 
from missing energy process $e N \to e N \phi$, the dashed purple line shows the  expected limit for projected statistics $\mbox{EOT}\simeq 5\times 10^{12}$.  
The solid blue line is the existing limit of the NA64e ($\mbox{EOT}\simeq 9.37\times 10^{11}$) due to the positron  annihilation mode $e^+e^- \to \phi$ with the primary $e^{-}$ beam,  the solid red line is the projected expected reach of NA64e for the positron annihilation channel with primary $e^+$ beam that corresponds to the $9.37\times 10^{11}$ positrons accumulated on target.
Dashed blue and red  lines correspond to the projected NA64e statistics $5\times10^{12}$ of the annihilation mode  $e^+ e^- \to \phi$ for the $e^-$ and $e^+$ primary beam, respectively.
Dashed orange line is the projected LDMX limits for $\mbox{EOT}\simeq  10^{15}$ from the $\phi$-strahlung process $e N \to e N \phi$. 
Dashed pink and yellow  lines correspond to the projected LDMX statistics $10^{15}$ of the annihilation mode $e^+ e^- \to \phi$ for the $e^-$ and $e^+$ primary beam, respectively.
Top Left, top right and bottom left panels correspond relic target for the Majorana, Dirac and 
scalar DM with parameters $\alpha_{\rm DM} = 0.5  \, (\mbox{GeV}^2)$
and $m_{\rm DM} / m_{\phi} = 1/3$.
The bottom right panel corresponds the relic target for the vector DM  with parameters $\alpha_{\rm V} = 0.5\, \mbox{GeV}^{-2}$ and $m_{\rm DM} / m_{\phi} = 2/5$.
}\label{fig:EP_ALL}
\end{figure*}

{ \it The light dark matter experiment (LDMX)}:
the LDMX is the projected electron fixed-target facility at Fermilab.
The schematic layout and the rules for the missing momentum event selection of the LDMX experiment are given in Ref.~\cite{Berlin:2018bsc}. 
The projected LDMX facility would employ the unique electron missing momentum technique~\cite{Mans:2017vej} that is complementary to the NA64$e$ facility.
The typical efficiencies $\eta_{\rm MED}^{\rm brem.}$ and $\eta_{\rm MED}^{ann.}$ are estimated to be at the level of $\simeq 50 \%$ for both electron~\cite{Akesson:2022vza} and positron beam options.
Thus, for the LDMX experiment we employ the following benchmark parameters  $(\rho~=~2.7~~\mbox{g cm}^{-3}$, $A~=~27~~\mbox{g mole}^{-1}$, $Z~=~13$, $x_{min}=0.7$, $X_0~=~8.9~\mbox{cm})$ and $L_T ~\simeq~0.4~X_0~\simeq~3.56~~\mbox{cm}$. The typical efficiencies $\eta_{\rm MED}^{\rm brem.}$ and $\eta_{\rm MED}^{ann.}$ are estimated to be at the level of $\simeq 50 \%$ for both electron~\cite{Akesson:2022vza} and positron beam options.
The energy of the primary beam is chosen to be $E_0\simeq16~~\mbox{GeV}$ and the projected moderate statistics corresponds to $\mbox{EOT}\simeq 10^{15}$ (it is planned however to collect $\mbox{EOT}\simeq 10^{16}$ by the final phase of experimental running after 2027, see e.~g.~Ref.~\cite{Akesson:2022vza} and references therein for detail).

\section{The resonant DM Production mechanism  \label{sec:ProductionMechanism}}

In this section we  provide a cross sections and widths of decay in case of scalar MED that are used for to  constrain the coupling between the scalar MED and the electron. In the following we use an notations for typical velocities and the inversed propagator squared:
\begin{align}
    &   \beta_{\rm DM}(s) \label{eq:defNoteBeta}
    = 
        \sqrt{1 - 4 m_{\rm  DM}^2 / s },
    \quad
        \beta_{e}(s) 
    = 
        \sqrt{1 - 4 m_{e}^2 / s },
\\
    &   D_{\rm  DM}^2(s)  \label{eq:defNoteD}
    = 
        (s - m_{\phi}^2)^2  
        +   
        m_{\phi}^2 (\Gamma^{tot}_{\rm  DM})^2.
\end{align}
For the calculation of both the decay width and cross section we employ the state-of-the-art FeynCalc  package~\cite{Shtabovenko:2020gxv,Shtabovenko:2016sxi} for the Wolfram Mathematica routine~\cite{Mathematica}.

As we discussed above, the scalar mediator $\phi$ can be produced in the bremsstrahlung-like reaction 
$e N \to e N \phi$ followed by the invisible decay $\phi \to \mbox{DM} + \mbox{DM}$.  
The details  of bremsstrahlung-like production of scalar MED in fixed-target experiments can be found 
elsewhere  in  Refs.~\cite{Voronchikhin:2023znz,Liu:2016mqv}.

The widths of decay of the scalar MED into vector \cite{Kaneta:2016wvf}, Dirac fermion 
\cite{Krnjaic:2015mbs,Marsicano:2018vin}, Majorana fermion \cite{Djouadi:2011aa}, scalar 
\cite{Djouadi:2011aa} DM, and electron-positron pair read, respectively, as follows:
\begin{align} 
    &   \Gamma_{\phi \to V V} \label{DecayWidthPhiToVV}
=
    \frac{1}{2}\cdot
    \frac{4 \pi \alpha_{V} m_\phi^3  }{32 \pi  } 
    \beta_{V}^3(m_{\phi}^2),
\\
    &     \Gamma_{\phi \to \psi \overline{\psi}} \label{DecayWidthPhiToPsiPsi} 
=
    \frac{4 \pi \alpha_\psi m_\phi }{8 \pi} 
    \beta_{\psi}^3(m_{\phi}^2),  
\\ 
    &     \Gamma_{\phi \to \chi \chi} \label{DecayWidthPhiToChiChi} 
=
    \frac{1}{2}\cdot
    \frac{4 \pi \alpha_\chi m_\phi }{8 \pi} 
    \beta_{\chi}^3(m_{\phi}^2),
\\ 
    &     \Gamma_{\phi \to S S} \label{DecayWidthPhiToSS} 
=
    \frac{1}{2}\cdot
    \frac{4 \pi \alpha_S  }{16 \pi m_\phi } 
    \beta_{S}(m_{\phi}^2), 
\\ 
    &     \Gamma_{\phi \to e^+e^-} \label{DecayWidthPhiToee} 
\simeq 
    \frac{(c^{\rm \phi}_{ee})^2 m_\phi }{8 \pi}
    \beta_{e}^3(m_{\phi}^2),   
\end{align}
where we use conventions defined by Eq.~(\ref{eq:defNoteBeta}). 
For the invisible mode we imply that  $m_{\rm MED} \gtrsim 2 m_{\text{\rm DM}}$ and 
$\Gamma_{{\rm MED} \to e^+e^-}~\ll~\Gamma_{\rm MED \to DM\;DM}$ throughout the paper and therefore it leads to the benchmark total decay widths  $\Gamma^{tot}_{\rm DM} \simeq \Gamma_{\rm MED \to DM\;DM}$.




The resonant total cross sections in case scalar,  Dirac, Majorana and vector 
DM read, respectively:
\begin{align}
    &  \sigma_{  e^+e^-  \to   \phi   \to S S } \label{eq:RTCSElectronToScalarDM}
= 
    \frac{1}{2}
    \frac{4 \pi \alpha_S (c^{\phi }_{ee})^2}{32 \pi }
    \frac{\beta_S(s) \beta_{e}(s)}{D_S^2(s)},
\\ 
    &   \sigma_{ e^+e^-  \to  \phi \to   \psi \overline{\psi} } \label{eq:RTCSElectronToDiracDM}
= 
    \frac{4 \pi \alpha_\psi (c^{\phi }_{ee})^2}{ 16 \pi }
    \frac{s \beta_\psi^3(s) \beta_{e}(s)}{D_\psi^2(s)},
\\
    &   \sigma_{ e^+e^-  \to  \phi \to   \chi \chi }\label{eq:RTCSElectronToMajoranaDM}
= 
    \frac{1}{2}
    \frac{4 \pi \alpha_\chi (c^{\phi }_{ee})^2}{ 16 \pi }
    \frac{s \beta_\chi^3(s) \beta_{e}(s)}{D_\chi^2(s)},
\\
    &   \sigma_{ e^+e^-  \to  \phi \to   V V }\label{eq:RTCSElectronToVectorDM}
= 
    \frac{1}{2}
    \frac{4 \pi \alpha_{V} (c^{\phi }_{ee})^2}{ 64 \pi }
    \frac{s^2 \beta_{V}^3(s) \beta_{e}(s)}{D_{V}^2(s)},
\end{align}
where the Mandelstam invariant is  $s  =  (p_{e^-} + p_{e^+})^2$.
In case of fixed-target experiments, the high-energy primary electron or positron beams are 
incident on the fixed dump that entails the development of an electromagnetic shower inside 
the thick active target. 
Thus, secondary positrons from the electromagnetic shower can annihilate on atomic electrons of target material via scalar MED into DM. 
One can see that typical momenta of the atomic electrons and secondary positrons are $p_{e^-} = (m_e, 0,0,0)$ and $p_{e^+} \simeq (E_{e^+}, 0, 0, E_{e^+})$, respectively, where are neglected the motion of atomic electrons. 
In addition, we focus only on ultrarelativistic positrons due to $m_e < m_{\rm DM}$ in the mass region of light DM, therefore,  $|\vect{p}_{e^+}| \simeq E_{e^+}$. As a result, the invariant mass $s$ takes the following expression:
\begin{equation}\label{eq:ExprSforPosinton}
    s  =  m_e^2 + 2 m_e E_{e^+} \simeq  2 m_e E_{e^+},
\end{equation}
that is employed in Eqs.~(\ref{eq:RTCSElectronToScalarDM}), (\ref{eq:RTCSElectronToDiracDM}), (\ref{eq:RTCSElectronToMajoranaDM}), (\ref{eq:RTCSElectronToVectorDM}), and (\ref{eq:constrCoulConst}) for the 
calculation  of the resonant missing energy signal yield.




It is important to mention that the resonant total cross section of process $f\overline{f} \to~R\to~l\overline{l}$ can be calculated by the Breit-Wigner (BW) resonant formula \cite{ParticleDataGroup:2022pth,Cheng:2023dau}:
\begin{multline}\label{eq:BWformiula}
    \sigma_{f\overline{f} \to~R\to~l\overline{l}}
\; =
    \frac{2 J + 1}{(2 S_1 \!+ \! 1)(2 S_2 \!+\! 1)}
\\
     \frac{16 \pi C_{BW} }{( 1 \! - \! 4 m^2_{f} / s)} 
    \frac{\Gamma_{ \rm in}(s)\Gamma_{ \rm out}(s)}{(s - m_R^2)^2 \! + \!  m_R^2(\Gamma^{tot}_{\rm  DM})^2},
\end{multline}
where  $\Gamma_{\rm tot}$ is a total width of decay of  
resonance,  $\Gamma_{ \rm in}$, $\Gamma_{ \rm out}$ are widths of decay of scalar MED in initial and final 
states, respectively, $J$ is a spins of resonance, $S_1$ and $S_2$ are spins of initial particles, $C_{BW} = 1$ 
and $C_{BW} = 2$  for different and identical particles in the initial state, respectively, that compensates the 
additional factor in the decay width in the case of identical particles \cite{Lvov:2018tgp,Cheng:2020iwk}.

By substituting the decay widths  
(\ref{DecayWidthPhiToVV}),
(\ref{DecayWidthPhiToPsiPsi}), (\ref{DecayWidthPhiToChiChi}), 
(\ref{DecayWidthPhiToSS}), and (\ref{DecayWidthPhiToee}) into 
the Breit-Wigner resonant formula (\ref{eq:BWformiula}) one can reproduce the result for the FeynCalc 
straightforward calculation of the  total cross sections 
(\ref{eq:RTCSElectronToScalarDM}), (\ref{eq:RTCSElectronToDiracDM}), 
(\ref{eq:RTCSElectronToMajoranaDM}), and  (\ref{eq:RTCSElectronToVectorDM}). 
That cross-check was used in order to validate our analytical calculations.

 \section{Thermal target \label{sec:FreezeOutMechanism}}
In this subsection, we briefly discuss cross sections of DM annihilation into electron-positron 
pair via the scalar MED  for different types of DM. We also plot regarding thermal target curves. 
In case of invisible decay mode the 
interaction can be realized via the   s-channel. 
The resonant total cross sections in the case of scalar MED and Dirac, Majorana, scalar, vector DM read,  respectively, as follows:
\begin{align}
    &   \sigma_{ \psi \overline{\psi} \! \to \!  \phi  \! \to \! e^+e^- }\label{eq:RTCSDiracDMtoElectron}
=
    \frac{4 \pi \alpha_\psi (c^{\phi }_{ee})^2}{16 \pi }
    \frac{s \beta_{\psi}(s) \beta_{e}^3(s)}{D_{\psi}^2(s)},
\\
    &   \sigma_{ \chi \chi \! \to \!  \phi  \! \to \! e^+e^- }\label{eq:RTCSMajoranaDMtoElectron}
=
    \frac{4 \pi \alpha_\chi (c^{\phi }_{ee})^2}{16 \pi }
    \frac{s \beta_{\chi}(s) \beta_{e}^3(s)}{D_{\chi}^2(s)},
\\
    &  \sigma_{ S S \! \to \!  \phi  \! \to \! e^+e^- }\label{eq:RTCSScalarDMtoElectron}
=
    \frac{4 \pi \alpha_S (c^{\phi }_{ee})^2}{8 \pi }
    \frac{\beta_{S}^{-1}(s) \beta_{e}^3(s)}{D_{S}^2(s)},
\\
    &  \sigma_{ V V \! \to \!  \phi  \! \to \! e^+e^- }\label{eq:RTCSVectorDMtoElectron}
=
    \frac{4 \pi \alpha_{V} (c^{\phi }_{ee})^2}{ 144 \pi }
    \frac{s^2 \beta_{V}(s) \beta_{e}^3(s)}{D_{V}^2(s)},
\end{align}
where we use the dark fine structure constant (\ref{eq:forCouplingConstant}) 
and~(\ref{eq:defNoteBeta}). 

In non-relativistic approximation, $s~\simeq~4 m^2_{\rm DM}$,  the typical product of  cross section  and velocity reads as follows: 
\begin{align}\label{eq:SimpleLowVelosityMajorana}
    &   v_{\rm Mol} \sigma_{ \chi \chi \! \to \!  \phi  \! \to \! e^+e^- }
\!=\! 
    \frac{  4 \pi \alpha_\chi (c^{\phi }_{ee})^2 m^2_{\chi} \beta_{e}^3(4 m^2_{\chi}) }
         {8 \pi D_{\chi}^2 \left(4 m^2_{\chi}\right)}
         v_{\rm Mol}^2,
\\ \label{eq:SimpleLowVelosityDirac}
    &   v_{\rm Mol} \sigma_{ \psi \overline{\psi} \!    \to \!  \phi  \! \to \! e^+e^- }
\!=\!
    \frac{  4 \pi \alpha_\psi (c^{\phi }_{ee})^2m^2_{\psi} \beta_{e}^3(4 m^2_{\psi}) }
         {8 \pi D_{\psi}^2 \left(4 m^2_{\psi}\right)}
         v_{\rm Mol}^2,
\\ \label{eq:SimpleLowVelosityScalar}
    &   v_{\rm Mol} \sigma_{ S S \! \to \!  \phi  \! \to \! e^+e^- }
\!=\! 
    \frac{  4 \pi \alpha_S (c^{\phi }_{ee})^2 \beta_{e}^3(4 m^2_{S})}
         {4 \pi D_{S}^2\left(4 m^2_{S}\right)},
\\ \label{eq:SimpleLowVelosityVector}
    &   v_{\rm Mol} \sigma_{ V V \! \to \!  \phi  \! \to \! e^+e^- }
\!=\! 
    \frac{  4 \pi \alpha_V (c^{\phi }_{ee})^2 m^4_{V} \beta_{e}^3(4 m^2_{V}) }
         {18 \pi D_{V}^2\left(4 m^2_{V}\right)}
         v_{\rm Mol}^2,
\end{align}
where the Moller's velocity  in the center of mass is defined by~\eqref{eq:MollerVelocityGeneral}.
One can see that  vector DM, Majorana and Dirac fermion DM annihilate in a p-wave, however the scalar DM
annihilates in a s-wave. As  an additional cross-check of 
the Eqs.~(\ref{eq:SimpleLowVelosityMajorana}),~(\ref{eq:SimpleLowVelosityDirac}), and~(\ref{eq:SimpleLowVelosityScalar}) we reproduce  
the expressions for Majorana DM \cite{Berlin:2018bsc,Djouadi:2011aa}, Dirac 
DM~\cite{Krnjaic:2015mbs,Buckley:2014fba} and scalar DM \cite{Djouadi:2011aa},  implying  
that  $D_{\rm DM}^2(4 m^2_{\rm DM})~\simeq~(4 m^2_{\rm DM} - m_{\phi}^2)^2$ 
in the non-relativistic approach. 


In Fig.~\ref{fig:RelAbAll} we show the thermal target curves  of different DM types that annihilates into $e^+e^-$ pair via the electron-specific scalar MED.
We compare the numerical calculation of the thermal target via  Eq.~(\ref{eq:GelminiTermAvCS}) and the analytical non-relativistic approximation defined by Eq.~(\ref{eq:LowVelocityExpan}). 

Remarkably that the numerical result for the thermal target curve  coincides fairly well with the 
analytical one  for the scalar DM that annihilates in a $s$-wave for the wide range of DM mass 
$1 \, \mbox{MeV}\lesssim  m_{S} \lesssim 10 \, \mbox{GeV}$ far from the resonant point 
$m_\phi \gtrsim 3 m_S$. It is worth noticing that for the fermionic DM (Dirac and Majorana), that  annihilates in a $p$-wave,
the discrepancy between analytical and numerical approach can be as 
large as $ \gtrsim 20~\%$ for the mass range of interest 
$1 \, \mbox{MeV}\lesssim  m_{\psi,\chi} \lesssim 10 \, \mbox{GeV}$ and  $m_\phi \gtrsim 3 m_{\psi, \chi}$.  
Moreover, the  difference between thermal target curve of Dirac DM and Majorana DM can be explained due to the  
particle-antiparticle coefficient $c$ (for details see e.~g. discussion after Eq.~(\ref{QandSdef1})). 

In order to 
cross-check our calculation, we reproduce the result of Ref.~\cite{Berlin:2018bsc} for the Majorana thermal target 
curve associated with the electron-specific scalar mediator. For vector DM the analytical approach of low-velocity regime breaks at small DM masses   $m_{V} \lesssim 10\, \mbox{MeV}$ and the regarding 
relative error can be relatively large for $m_\phi \simeq 3 m_V$. This  discrepancy is mitigated in 
the region far from the resonant  point  $m_\phi \simeq 6 m_V$. 
In the following section, we rely on numerical approach for vector thermal target 
calculation~\cite{Gondolo:1990dk,Wells:1994qy,Choi:2017mkk}.

\section{The experimental reach \label{sec:ExpectedReach}}

In this section we discuss current and expected experimental reach of the fixed-target facilities as NA64e and LDMX 
for both primary electron and positron  beams, implying that scalar mediator decays into the invisible  mode.
We use the typical number of signal events $N_{\rm sign.} \gtrsim  2.3$ and obtain the $90\% \mbox{~C.~L.}$ 
exclusion limit on the coupling constant between the electron and the scalar MED. 
Also, the background free case and the null-results of the missing energy events for both NA64e and LDMX 
experiments are used.
Thus, for the signal yield that originates from MED-strahlung and $e^+e^-$~annihilation into DM we put 
$N_{\rm sign. }\simeq N^{\rm brem. }_{\mbox{\scriptsize MED}}+N^{\rm ann. }_{\mbox{\scriptsize MED}}$.
In Fig.~\ref{fig:EP_ALL} we show the experimental reach of the NA64e and LDMX experiments. 

We use the current statistics of NA64e as ${\rm EOT}\simeq9.37\times10^{11}$ and the projected statistics for NA64e and LDMX as ${\rm EOT}\simeq5\times10^{12}$ and ${\rm EOT}\simeq10^{15}$, respectively. 
 We assume same statistics for both positron and electron primary beams in case of calculation of 
 positron annihilation mode. 
We note that, for the  $e^+e^-$ annihilation mode,  the projected positron beam provides a more 
stringent constraint on the electron-MED coupling than the electron primary beam bounds  due to the 
presence of positrons in the first generation of the electromagnetic shower in the 
active target~\cite{Voronchikhin:2023znz}.

Remarkably, the NA64e experiment rules out the  Dirac and vector DM for the currently 
accumulated  statistics ${\rm EOT}\simeq9.37\times10^{11}$. In addition, the NA64e resonant 
enhancement  of the exclusion limit in the mass range $0.23 \, \mbox{GeV} \lesssim m_{\phi} \lesssim 0.32\, \mbox{GeV}$ and the bounds from  $\phi$-strahlung around  $10^{-2}\, \mbox{GeV} \lesssim m_\phi \lesssim 0.1 \, \mbox{GeV}$ allow to exclude the Majorana DM.  However, for the 
scalar DM the NA64e rules out only small part of relic thermal target parameter space below 
$m_{\phi} \lesssim 5\times 10^{-3}\, \mbox{GeV}$.  Note that the LDMX can significantly constrain the electron-MED coupling due to the sufficiently large 
planned accumulated statistics.

\section{Conclusion
\label{sec:Conclusion}}
In the present paper, we studied a  sub-GeV range of simplified 
DM  scenarios associated with the electron-specific scalar mediator 
and evaluated the sensitivity of electron fixed-target experiments such as NA64e and LDMX. 
In particular, we derived  new constraints on the coupling of electrophilic scalar MED in case 
of $e^+ e^-$ annihilation mode and showed that these complement the bounds  
from the widely exploited $\phi$-strahlung production, $e N \to e N \phi$, followed by 
the invisible decay, $\phi \to \mbox{DM}+\mbox{DM}$.
In particular, the updated reach of NA64e and LDMX  can be pushed down by a factor of 
$\mathcal{O}(1)$ for the specific range of  light DM mass. 
We studied the different mechanism of DM production for the regarding experiments and showed 
that the resonant annihilation mode $e^+ e^- \to \phi \to {\rm DM} + {\rm DM}$ can be a viable 
mechanism of the electron-specific MED production that provides an improved constraints for   
Majorana, Dirac, scalar and vector types of DM.
We showed that NA64e rules out the typical parameter space of Majorana, Dirac and vector DM
for the specific set of benchmark parameters.  
\begin{acknowledgments} 
We would like to thank A.~Celentano,  P.~Crivelli, S.~Demidov, R.~Dusaev,  S.~Gninenko, D.~Gorbunov,  
M.~Kirsanov, N.~Krasnikov, V.~Lyubovitskij, L.~Molina Bueno,  A.~Pukhov, A. Shevelev, H.~Sieber, and 
A.~Zhevlakov    for very helpful discussions and  correspondences.
This work was supported by The Ministry of Science and Higher Education of the Russian Federation in part of the Science program (Project № FSWW-2023-0003)
\end{acknowledgments}

\bibliography{bibl}

\end{document}